\documentclass[10pt,letterpaper,twocolumn,english,journal]{IEEEtran}
\usepackage[T1]{fontenc}
\usepackage{color}
\usepackage{float}
\usepackage{amsthm}
\usepackage{amsmath}
\usepackage{amssymb}
\usepackage{stackrel}
\usepackage{graphicx}
\usepackage{esint}

\makeatletter

\pdfpageheight\paperheight
\pdfpagewidth\paperwidth

\floatstyle{ruled}
\newfloat{algorithm}{tbp}{loa}
\providecommand{\algorithmname}{Algorithm}
\floatname{algorithm}{\protect\algorithmname}

\theoremstyle{plain}
\newtheorem{thm}{\protect\theoremname}
\theoremstyle{plain}
\newtheorem{lem}[thm]{\protect\lemmaname}

\usepackage{cite}\usepackage{algorithm}\usepackage{algorithmicx}\usepackage{algpseudocode}\usepackage{babel}

\providecommand{\theoremname}{Theorem}

\allowdisplaybreaks \allowdisplaybreaks[4]
\usepackage{babel}

\makeatother

\usepackage{babel}
\providecommand{\lemmaname}{Lemma}
\providecommand{\theoremname}{Theorem}

\begin{document}

\title{A Unified Bayesian Inference Framework for Generalized Linear Models}

\author{\IEEEauthorblockN{Xiangming Meng, Sheng Wu, and Jiang Zhu}%
\thanks{X. Meng is with Huawei Technologies, Co. Ltd., Shanghai, China. (e-mail:
\mbox{%
mengxm11%
}@gmail.com). S. Wu is with Tsinghua Space Center, Tsinghua Univsersity,
Beijing, China. (e-mail: %
\mbox{%
thuraya%
}@tsinghua.edu.cn). J. Zhu is with Ocean College, Zhejiang University,
Zhoushan, China. (e-mail: %
\mbox{%
jiangzhu16%
}@zju.edu.cn). %
}}
\maketitle
\begin{abstract}
In this letter, we present a unified Bayesian inference framework
for generalized linear models (GLM) which iteratively reduces the
GLM problem to a sequence of standard linear model (SLM) problems.
This framework provides new perspectives on some established GLM algorithms
derived from SLM ones and also suggests novel extensions for some
other SLM algorithms. Specific instances elucidated under such framework
are the GLM versions of approximate message passing (AMP), vector
AMP (VAMP), and sparse Bayesian learning (SBL). It is proved that
the resultant GLM version of AMP is equivalent to the well-known generalized
approximate message passing (GAMP). Numerical results for 1-bit quantized
compressed sensing (CS) demonstrate the effectiveness of this unified
framework.\end{abstract}


\section{Introduction}

The approximate message passing (AMP) algorithm \cite{donoho2009message},
first proposed by Donoho \textit{et al. }in the field of compressed
sensing (CS)\cite{Candes-Introduction-to-CS}, is one state-of-the-art
algorithm for Bayesian inference over the standard linear model (SLM)\cite{donoho2010message,krzakala2012probabilistic,Som2012,vila2013expectation,ziniel2013dynamic,ziniel2013efficient,wu2014-JSTSP,meng2015expectation,wang2015multiuser}.
\textcolor{black}{Since its first publication, much effort has been
done to extend AMP\cite{Ma2014Turbo,Rangan2016Vector,Schniter2016Vector,Schniter2016Denoising,Xue2016D,Ma2016Orthogonal,Liu2016Generalized,He2017Generalized}.
In \cite{Rangan2016Vector}, vector AMP (VAMP) was proposed by extending
the measurement matrix from i.i.d. sub Gaussian to right-rotationally
invariant matrix. For partial discrete Fourier transform (DFT) matrix,
a variant of AMP was proposed \cite{Ma2014Turbo}. For unitarily-invariant
matrices, the authors in \cite{Ma2016Orthogonal} proposed orthogonal
AMP (OAMP) using de-correlated linear estimator and divergence-free
nonlinear estimator. }

\textcolor{black}{However, in some applications such as quantized
CS \cite{Kamilov2012Message}, linear classification \cite{bishop2006pattern},
phase retrieval \cite{Jaganathan2015Phase}, etc., the measurements
are nonlinear transform of the input signal, to which SLM no longer
applies. To this end, Rangan extended AMP to generalized approximate
message passing (GAMP) \cite{rangan2011generalized,Rangan2012GAMP}
for generalized linear models (GLM) so that it can efficiently recover
the signal of interest from nonlinear measurements. Recently, both
VAMP \cite{Rangan2016Vector} and turbo CS \cite{Ma2014Turbo} have
been extended to handle GLM problems \cite{Schniter2016Vector,Liu2016Generalized,He2017Generalized}.}

In this letter, a unified Bayesian inference framework for GLM inference
is presented whereby the original GLM problem is iteratively reduced
to a sequence of SLM problems. Under such framework, a variety of
inference methods for SLM can be easily extended to GLM. As specific
instances, we extend AMP, VAMP, and sparse Bayesian learning (SBL)
\cite{tipping2001sparse,wipf2004sparse} to GLM under this framework
and obtain three GLM inference algorithms termed as Gr-AMP, Gr-VAMP,
and Gr-SBL, respectively. It is proved that Gr-AMP is equivalent to
GAMP \cite{rangan2011generalized}. Note that although both AMP and
VAMP have been extended to GLM in \cite{rangan2011generalized} and
\cite{Schniter2016Vector}, each of them was derived from a different
perspective tailored to one specific SLM algorithm only. Under the
proposed framework, however, we unveil that they actually share a
common rule and more importantly, this rule also suggests novel extensions
for some other SLM algorithms, e.g., SBL. Thus the main contribution
of this letter is providing a simple and unified Bayesian framework
for the understanding and extension of SLM algorithms to GLM. Numerical
results for 1-bit quantized CS demonstrate the effectiveness of this
framework.

\section{Problem Description}

\begin{figure}
\centering{}\includegraphics[width=8.5cm]{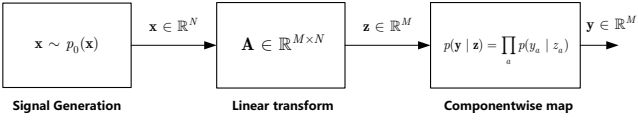}\protect\caption{Generalized linear models (GLM)\cite{Rangan2012GAMP}.}
\label{GLM_model}
\end{figure}

Consider the system model of GLM shown in Fig. \ref{GLM_model}. The
signal $\mathbf{x}\in\mathbb{R}^{N}$ is generated following a prior
distribution $p_{0}\bigl(\mathbf{x}\bigr)$. Then, $\mathbf{x}$ passes
through a linear transform $\mathbf{z}=\mathbf{Ax}$, where $\mathbf{A}\in\mathbb{R}^{M\times N}$
is a known matrix and the measurement ratio is $\delta=M/N$. The
observation vector $\mathbf{y}$ is obtained through a component-wise
random mapping, which is described by a factorized conditional distribution
\begin{equation}
p\bigl(\mathbf{y}|\boldsymbol{\mathbf{z}}\bigr)=\stackrel[a=1]{M}{\prod}p\bigl(y_{a}|z_{a}\bigr)=\stackrel[a=1]{M}{\prod}p\bigl(y_{a}|z_{a}=\sum_{i=1}^{N}A_{ai}x_{i}\bigr).\label{eq:ouput_map}
\end{equation}

The goal of GLM inference is to compute the minimum mean square error
(MMSE) estimate of $\mathbf{x}$, i.e., $\mathsf{E}(\mathbf{x}|\mathbf{y})$,
where the expectation is taken with respect to (w.r.t.) the posterior
distribution $p\bigl(\mathbf{x}|\mathbf{y}\bigr)\propto p_{0}\bigl(\mathbf{x}\bigr)p\bigl(\mathbf{y}|\mathbf{z}=\mathbf{Ax}\bigr)$,
where $\propto$ denotes identity up to a normalization constant.
In this letter, $p_{0}\bigl(\mathbf{x}\bigr)$ and $p\bigl(\mathbf{y}|\boldsymbol{\mathbf{z}}\bigr)$
are assumed known. Extensions to scenarios with unknown $p_{0}\bigl(\mathbf{x}\bigr)$
or $p\bigl(\mathbf{y}|\boldsymbol{\mathbf{z}}\bigr)$ are possible
\cite{vila2013expectation,krzakala2012probabilistic,Marttinen2013Approximate}.
Note that the well-known SLM is a special case of GLM which reads
\begin{equation}
\mathbf{y}=\mathbf{Ax}+\mathbf{w},\label{eq:linear_model}
\end{equation}
where $\mathbf{w}$ is a Gaussian noise vector with mean $\mathbf{0}$
and covariance matrix $\sigma^{2}\mathbf{I}_{M}$, i.e., $\mathbf{w}\sim\mathcal{N}\bigl(\mathbf{w};\mathbf{\mathbf{0}},\sigma^{2}\mathbf{I}_{M}\bigr),$
where $\mathbf{I}_{M}$ denotes identity matrix of dimension $M$.

\section{\label{sec:Inference-Framework-for}Unified Inference Framework}

\begin{figure}
\centering{}\includegraphics[width=8.5cm]{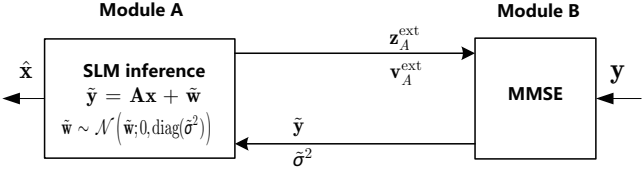}\protect\caption{A unified inference framework for GLM.}
\label{Turbo_framework}
\end{figure}

In this section we present a unified Bayesian inference framework,
shown in Fig. \ref{Turbo_framework}, for the GLM problem.\textcolor{red}{{}
}It consists of two modules: an SLM inference module A and a component-wise
MMSE module B. The messages $\mathbf{z}_{\textrm{A}}^{\textrm{ext}}$
and $\mathbf{v}_{\textrm{A}}^{\textrm{ext}}$ (defined in (\ref{eq:mean_z_ext}),
(\ref{eq:var_z_ext})) form the outputs of module A and inputs to
module B, while the messages $\mathbf{\tilde{y}}$ and $\tilde{\boldsymbol{\sigma}}^{2}$
(defined in (\ref{eq:mean_z_ext_B2A}), (\ref{eq:var_z_ext_B2A}))
form the outputs of module B and inputs to module A. To perform GLM
inference, we alternate between the two modules in a turbo manner\cite{berrou1996near}.
Assuming the prior distribution of $\mathbf{z}$ as Gaussian with
mean $\mathbf{z}_{\textrm{A}}^{\textrm{ext}}$ and variance $\mathbf{v}_{\textrm{A}}^{\textrm{ext}}$,
module B performs component-wise MMSE estimate and outputs $\mathbf{\tilde{y}}$
and $\tilde{\boldsymbol{\sigma}}^{2}$. Module A then performs SLM
inference over a pseudo-SLM $\mathbf{\tilde{y}}=\mathbf{Ax}+\mathbf{\tilde{w}},$
$\mathbf{\tilde{w}}\sim\mathcal{N}\bigl(\mathbf{\tilde{w}};\mathbf{\mathbf{0}},\mathsf{diag}(\tilde{\boldsymbol{\sigma}}^{2})\bigr)$
and updates the outputs $\mathbf{z}_{\textrm{A}}^{\textrm{ext}}$
and $\mathbf{v}_{\textrm{A}}^{\textrm{ext}}$ to module B. This process
continues iteratively until convergence or some predefined stopping
criteria is satisfied, where one iteration refers to one pass of messages
from $\textrm{B}$ to $\textrm{A}$ and then back to $\textrm{B}$.

Specifically, in the $t$-th iteration, the inputs $z_{a,A}^{\textrm{ext}}(t-1)$
and $v_{a,A}^{\textrm{ext}}(t-1)$ to module B are treated as the
prior mean and variance of $z_{a}$, respectively. Assuming the prior
is Gaussian, the posterior mean and variance of $z_{a}$ can be computed
in module B by component-wise MMSE estimate, i.e.,
\begin{align}
z_{a,B}^{\textrm{post}}(t) & =\mathsf{E}(z_{a}|z_{a,A}^{\textrm{ext}}(t-1),v_{a,A}^{\textrm{ext}}(t-1)),\label{eq:Z_post_B}\\
v_{a,B}^{\textrm{post}}(t) & =\mathsf{Var}(z_{a}|z_{a,A}^{\textrm{ext}}(t-1),v_{a,A}^{\textrm{ext}}(t-1)),\label{eq:Vz_post_B}
\end{align}
where the expectation is taken w.r.t. the posterior distribution
\begin{equation}
q_{z}^{t-1}\left(z_{a}\right)\propto p\bigl(y_{a}|z_{a}\bigr)\mathcal{N}\bigl(z_{a};z_{a,A}^{\textrm{ext}}(t-1),v_{a,A}^{\textrm{ext}}(t-1)\bigr).\label{eq:post_prob_xi-1-1}
\end{equation}
According to the turbo principle\textcolor{black}{{} \cite{berrou1996near}},
the posterior mean and variance cannot be directly used as the output
message since they incorporate information of the input messages.
Instead, the so-called extrinsic mean $\tilde{y}_{a}(t)$ and variance
$\tilde{\sigma}_{a}^{2}(t)$ of $z_{a}$ are calculated by excluding
the contribution of input messages $z_{a,A}^{\textrm{ext}}(t-1)$
and $v_{a,A}^{\textrm{ext}}(t-1)$, which are given as\textcolor{black}{{}
\cite{Guo2011}} 
\begin{alignat}{1}
\tilde{\sigma}_{a}^{2}(t) & =\Bigl(\frac{1}{v_{a,B}^{\textrm{post}}(t)}-\frac{1}{v_{a,A}^{\textrm{ext}}(t-1)}\Bigr)^{-1},\label{eq:var_z_ext_B2A}\\
\tilde{y}_{a}(t) & =\tilde{\sigma}_{a}^{2}(t)\Bigl(\frac{z_{a,B}^{\textrm{post}}(t)}{v_{a,B}^{\textrm{post}}(t)}-\frac{z_{a,A}^{\textrm{ext}}(t-1)}{v_{a,A}^{\textrm{ext}}(t-1)}\Bigr).\label{eq:mean_z_ext_B2A}
\end{alignat}
For more details of the turbo principle and extrinsic information,
please refer to \textcolor{black}{\cite{berrou1996near,wang1999iterative}.
Then in module A, similar as \cite{wang1999iterative}, $\tilde{y}_{a}(t)$
can be viewed as the pseudo-observation of an equivalent additive
Gaussian channel with $z_{a}$ as input, i.e., 
\begin{equation}
\tilde{y}_{a}(t)=z_{a}+\tilde{w}_{a}(t),\label{eq:Equivalent_observation_model}
\end{equation}
where $\tilde{w}_{a}(t)\sim\mathcal{N}\bigl(\tilde{w}_{a}(t);0,\tilde{\sigma}_{a}^{2}(t)\bigr)$.
In matrix form, (\ref{eq:Equivalent_observation_model}) can be rewritten
as an SLM form}
\begin{equation}
\mathbf{\tilde{y}}(t)=\mathbf{Ax}+\mathbf{\tilde{w}}(t),\label{eq:Psudo_SLM_model}
\end{equation}
where $\mathbf{\tilde{w}}(t)\sim\mathcal{N}\bigl(\mathbf{\tilde{w}}(t);\mathbf{0},\mathsf{diag}(\tilde{\boldsymbol{\sigma}}^{2}(t))\bigr)$.
\textcolor{black}{The underlying theme in (\ref{eq:Equivalent_observation_model})
is to }approximate the estimation error ($\tilde{y}_{a}(t)-z_{a}$)
as Gaussian noise whose mean is zero and variance is the extrinsic
variance $\tilde{\sigma}_{a}^{2}(t)$. \textcolor{black}{Such idea
was first used (to our knowledge) in \cite{wang1999iterative} (formula
(47) in \cite{wang1999iterative}). Rigorous theoretical analysis
on this will be future work.}

\textcolor{black}{As a result, the original GLM problem reduces to
an SLM problem so that one can estimate $\mathbf{x}$ by performing
SLM inference over (\ref{eq:Psudo_SLM_model}) to obtain the posterior
estimates. Specially, if GLM itself reduces to SLM, then from (\ref{eq:var_z_ext_B2A})
and (\ref{eq:mean_z_ext_B2A}), we obtain $\tilde{y}_{a}(t)=y_{a}$,
$\tilde{\sigma}_{a}^{2}(t)=\sigma^{2}$, which is exactly the original
SLM problem. Given the posterior mean and variance estimates of $x_{i}$,
the posterior mean $z_{a,A}^{\textrm{post}}(t)$ and variance $v_{a,A}^{\textrm{post}}(t)$
of $z_{a}$ can be obtained using the constraint $\mathbf{z}=\mathbf{Ax}$.
Note that the specific method to calculate $z_{a,A}^{\textrm{post}}(t)$
and $v_{a,A}^{\textrm{post}}(t)$ may vary for different SLM inference
methods, as shown in Section \ref{sec:GAMP-under-the}. Given $z_{a,A}^{\textrm{post}}(t)$
and $v_{a,A}^{\textrm{post}}(t)$, the extrinsic mean $z_{a,A}^{\textrm{ext}}(t)$
and variance $v_{a,A}^{\textrm{ext}}(t)$ of $z_{a}$ can be computed
as \cite{Guo2011}}
\begin{alignat}{1}
v_{a,A}^{\textrm{ext}}(t) & =\Bigl(\frac{1}{v_{a,A}^{\textrm{post}}(t)}-\frac{1}{\tilde{\sigma}_{a}^{2}(t)}\Bigr)^{-1},\label{eq:var_z_ext}\\
z_{a,A}^{\textrm{ext}}(t) & =v_{a,A}^{\textrm{ext}}(t)\Bigl(\frac{z_{a,A}^{\textrm{post}}(t)}{v_{a,A}^{\textrm{post}}(t)}-\frac{\tilde{y}_{a}(t)}{\tilde{\sigma}_{a}^{2}(t)}\Bigr),\label{eq:mean_z_ext}
\end{alignat}
which then form the input messages to module B in the next iteration.
Note that (\ref{eq:var_z_ext}) - (\ref{eq:mean_z_ext}) and (\ref{eq:var_z_ext_B2A})
- (\ref{eq:mean_z_ext_B2A}) apply the same rule but are different
operations in different modules. 

\textcolor{black}{This process continues until convergence or some
predefined stopping criteria is satisfied.} The corresponding algorithm
framework is shown in Algorithm \ref{Algorithm_GLM_framework}. \textcolor{black}{Note
that in step 3), one or more iterations can be} performed for iterative
SLM methods. Under such framework, a variety of inference methods
for SLM can be easily extended to GLM. As specific instances, we will
show in Section \ref{sec:GAMP-under-the} how AMP, VAMP, and SBL can
be extended to GLM within this framework. 

\begin{algorithm}
\protect\caption{Unified framework for GLM inference}

\begin{raggedright}
1) Initialization: $\mathbf{z}_{A}^{\textrm{ext}}(0),\mathbf{v}_{A}^{\textrm{ext}}(0)$,
$t=1$;
\par\end{raggedright}

2) Update $\tilde{\boldsymbol{\sigma}}^{2}(t)$, $\mathbf{\tilde{y}}(t)$
as (\ref{eq:var_z_ext_B2A}) and (\ref{eq:mean_z_ext_B2A});

3) Perform SLM inference over $\mathbf{\tilde{y}}(t)=\mathbf{Ax}+\mathbf{\tilde{w}}(t)$,
$\mathbf{\tilde{w}}(t)\sim\mathcal{N}\bigl(\mathbf{\tilde{w}}(t);0,\mathsf{diag}(\tilde{\boldsymbol{\sigma}}^{2}(t))\bigr)$
via one SLM inference method;

4) Update $\mathbf{v}_{A}^{\textrm{ext}}(t)$, $\mathbf{z}_{A}^{\textrm{ext}}(t)$
as (\ref{eq:var_z_ext}) and (\ref{eq:mean_z_ext});

5) Set $t\leftarrow t+1$ and proceed to step 2) until $t>T_{max}$.
\label{Algorithm_GLM_framework}
\end{algorithm}

\section{\label{sec:GAMP-under-the}Extending AMP, VAMP, and SBL to GLM}

\subsection{From AMP to Gr-AMP and GAMP}

\begin{figure}
\begin{centering}
\includegraphics[scale=0.32]{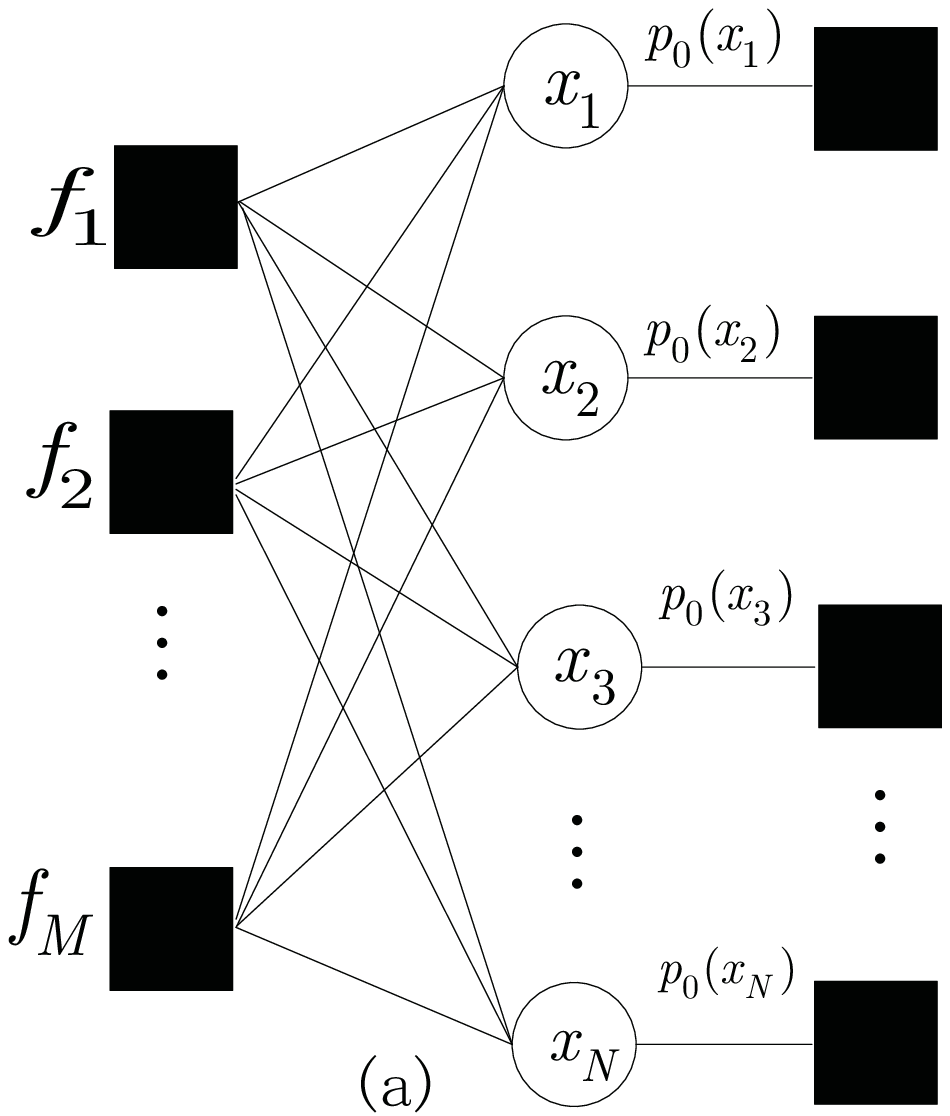} $\qquad$\includegraphics[scale=0.32]{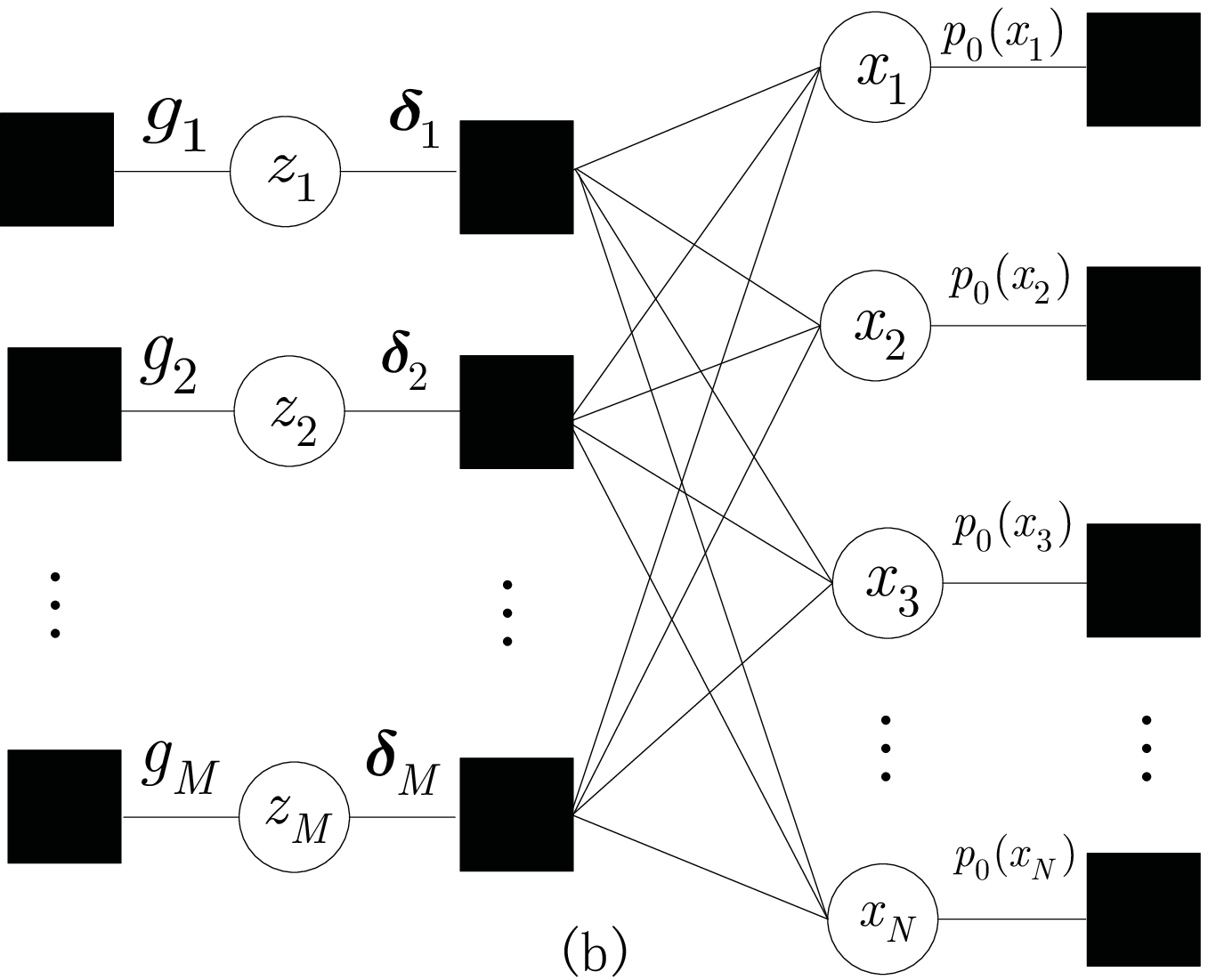}
\par\end{centering}

\protect\caption{Two equivalent factor graph representations of SLM. }
\label{fig:Factor Graph} 
\end{figure}

First review AMP for SLM. Many variants of AMP exist and in this letter
we adopt the Bayesian version in \cite{rangan2011generalized} and
\cite{krzakala2012probabilistic}. Assuming the prior distribution
$p_{0}\bigl(\mathbf{x}\bigr)=\prod_{i}p_{0}(x_{i})$, the factor graph
for SLM (\ref{eq:linear_model}) is shown in Fig. \ref{fig:Factor Graph}
(a), where $f_{a}$ denotes the Gaussian distribution $\mathcal{N}\bigl(y_{a};\sum_{i=1}^{N}A_{ai}x_{i},\sigma^{2}\bigr)$.
Denote by $m_{i\rightarrow f_{a}}^{t}(x_{i})=p_{0}(x_{i})\prod_{b\neq a}m_{f_{b}\rightarrow x_{i}}^{t-1}(x_{i})$
the message from $x_{i}$ to $f_{a}$ in the $t$-th iteration, with
mean and variance of $x_{i}$ being $\hat{x}_{i\rightarrow a}^{t}$
and $\tau_{i\rightarrow a}^{t}$. Then the message from $f_{a}$ to
$x_{i}$ is $m_{f_{a}\rightarrow x_{i}}^{t}(x_{i})=\int\mathcal{N}\bigl(y_{a};\sum_{i=1}^{N}A_{ai}x_{i},\sigma^{2}\bigr)\prod_{j\neq i}m_{j\rightarrow f_{a}}^{t}(x_{j})dx_{j}$,
with mean and variance of $x_{i}$ being $\hat{x}_{a\rightarrow i}^{t}$
and $\tau_{a\rightarrow i}^{t}$, respectively. Define 
\begin{align}
\Sigma_{i}(t) & \triangleq\bigl(\sum_{a}\frac{1}{\tau_{a\rightarrow i}^{t}}\bigr)^{-1},r_{i}(t)\triangleq\Sigma_{i}(t)\bigl(\sum_{a}\frac{\hat{x}_{a\rightarrow i}^{t}}{\tau_{a\rightarrow i}^{t}}\bigr)^{-1},\label{eq:Vi_Ri_def}\\
Z_{a}(t) & \triangleq\sum_{i}A_{ai}\hat{x}_{i\rightarrow a}^{t},\;\; V_{a}(t)\triangleq\sum_{i}A_{ai}^{2}\tau_{i\rightarrow a}^{t}.\label{eq:Za_Va_def}
\end{align}
By the central limit theorem and neglecting high order terms, the
resultant AMP is shown in Algorithm \ref{CAMP}. Note that the approximation
in AMP is asymptotically exact in large system limit, i.e., $N,M\rightarrow\infty$
with $M/N\rightarrow\delta$ \cite{rangan2011generalized,krzakala2012probabilistic}.
The expectation operation in the posterior mean $\mathsf{E}(x_{i}|r_{i}(t-1),\Sigma_{i}(t-1))$
and variance $\mathsf{Var}(x_{i}|r_{i}(t-1),\Sigma_{i}(t-1))$ in
line (D3) and (D4) are w.r.t. the posterior distribution $q_{x}^{t}\left(x_{i}\right)\propto p_{0}(x_{i})\mathcal{N}\bigl(x_{i};r_{i}(t),\Sigma_{i}(t)\bigr).$
For more details, refer to \cite{rangan2011generalized,krzakala2012probabilistic}.

\begin{algorithm}
\protect\caption{AMP Algorithm \cite{rangan2011generalized}\cite{krzakala2012probabilistic}}

\begin{raggedright}
1) Initialization: $\hat{x}_{i}(0)$,$V_{a}(0),Z_{a}(0)$, $t=1$;
\par\end{raggedright}

2) Variable node update: For $i=1,...,N$
\begin{alignat*}{1}
{\scriptstyle \Sigma_{i}(t-1)} & {\scriptstyle =\Bigl[\sum_{a}\frac{A_{ai}^{2}}{\sigma^{2}+V_{a}(t-1)}\Bigr]^{-1},}\qquad\qquad\quad\;\qquad\quad\;\;(\textrm{D}1)\\
{\scriptstyle r_{i}(t-1)} & {\scriptstyle =\hat{x}_{i}(t-1)+\Sigma_{i}(t-1)\sum_{a}\frac{A_{ai}\bigl(y_{a}-Z_{a}(t-1)\bigr)}{\sigma^{2}+V_{a}(t-1)},}\qquad\,(\textrm{D}2)\\
{\scriptstyle \hat{x}_{i}(t)} & {\scriptstyle =\mathsf{E}(x_{i}|r_{i}(t-1),\Sigma_{i}(t-1)),}\:\qquad\qquad\qquad\qquad\,(\textrm{D}3)\\
{\scriptstyle \tau_{i}(t)} & {\scriptstyle =\mathsf{Var}(x_{i}|r_{i}(t-1),\Sigma_{i}(t-1)).}\:\qquad\qquad\qquad\quad\;\,(\textrm{D}4)
\end{alignat*}

3) Factor node update: For $a=1,\ldots,M$
\begin{align*}
{\displaystyle {\scriptstyle V_{a}(t)}} & {\scriptstyle =\sum_{i}A_{ai}^{2}\tau_{i}(t),}{\textstyle \textrm{ }}\qquad\qquad\qquad\qquad\quad\;\;\;\qquad\,\;(\textrm{D}5)\\
{\scriptstyle Z_{a}(t)} & {\scriptscriptstyle =}{\scriptstyle \sum_{i}A_{ai}\hat{x}_{i}(t)-\frac{V_{a}(t)\bigl(y_{a}-Z_{a}(t-1)\bigr)}{\sigma^{2}+V_{a}(t-1)}.}\qquad\;\;\,\,\qquad\;(\textrm{D}6)
\end{align*}

4) Set $t\leftarrow t+1$ and proceed to step 2) until $t>\textrm{Iter}{}_{SLM}$.
\label{CAMP}
\end{algorithm}

Then we show how to extend AMP to GLM under the unified framework
\textcolor{black}{in Section \ref{sec:Inference-Framework-for}}.
As shown in Algorithm \ref{Algorithm_GLM_framework}, the key lies
in calculating the extrinsic mean $\mathbf{z}_{A}^{\textrm{ext}}$
and extrinsic variance $\mathbf{v}_{A}^{\textrm{ext}}$. For AMP,
fortunately, these two values have already been calculated, as shown
in Lemma 1. 
\begin{lem}
After performing AMP in module A, the output extrinsic mean $\mathbf{z}_{A}^{\textrm{ext}}$
and variance $\mathbf{v}_{A}^{\textrm{ext}}$ to module B can be computed
as 
\begin{align}
z_{a,A}^{\textrm{ext}}(t)=Z_{a}(t),\quad v_{a,A}^{\textrm{ext}}(t)=V_{a}(t) & ,\label{eq:V_z_Ext_A2B}
\end{align}
where $Z_{a}(t)$, $V_{a}(t)$ are the results of AMP in line (D6)
and (D5) of Algorithm \ref{CAMP}, respectively.\end{lem}
\begin{IEEEproof}
From (\ref{eq:var_z_ext}) and (\ref{eq:mean_z_ext}), the posterior
mean $z_{a,A}^{\textrm{post}}(t)$ and variance $v_{a,A}^{\textrm{post}}(t)$
of $z_{a}$ are needed to calculate $z_{a,A}^{\textrm{ext}}(t)$ and
$v_{a,A}^{\textrm{ext}}(t)$. However, in original AMP, they are not
explicitly given. To this end, we present an alternative factor graph
for SLM in Fig. \ref{fig:Factor Graph} (b), where $\delta_{a}$ denotes
the Dirac function $\delta\bigl(z_{a}-\sum_{i=1}^{N}A_{ai}x_{i}\bigr)$
and $g_{a}$ denotes Gaussian distribution $\mathcal{N}\bigl(z_{a};\tilde{y}_{a}(t),\tilde{\sigma}_{a}^{2}(t)\bigr)$.
The two factor graph representations in Fig. \ref{fig:Factor Graph}
are equivalent so that the message from $x_{i}$ to $\delta_{a}$
is precisely $m_{i\rightarrow\delta_{a}}^{t}(x_{i})=m_{i\rightarrow f_{a}}^{t}(x_{i})$.
Thus, the message from $\delta_{a}$ to $z_{a}$ is $m_{\delta_{a}\rightarrow z_{a}}^{t}(z_{a})=\int\delta\bigl(z_{a}-\sum_{i=1}^{N}A_{ai}x_{i}\bigr)\prod_{i}m_{i\rightarrow f_{a}}^{t}(x_{i})dx_{i}.$
Recall the definitions of $Z_{a}(t)$ and $V_{a}(t)$ in (\ref{eq:Za_Va_def}).
As in AMP, according to the central limit theorem, $z_{a}=\sum_{i}A_{ai}x_{i}$
can be approximated as Gaussian random variable whose mean and variance
are $Z_{a}(t)$ and $V_{a}(t)$, i.e., $m_{\delta_{a}\rightarrow z_{a}}^{t}(z_{a})=\mathcal{N}\bigl(z_{a};Z_{a}(t),V_{a}(t)\bigr)$.
Since the message from $g_{a}$ to $z_{a}$ is $m_{g_{a}\rightarrow z_{a}}^{t}(z_{a})=\mathcal{N}\bigl(z_{a};\tilde{y}_{a}(t),\tilde{\sigma}_{a}^{2}(t)\bigr)$,
the posterior distribution of $z_{a}$ can be calculated as $q^{t}(z_{a})\propto m_{\delta_{a}\rightarrow z_{a}}^{t}(z_{a})m_{g_{a}\rightarrow z_{a}}^{t}(z_{a})\propto\mathcal{N}\bigl(z_{a};z_{a,A}^{\textrm{post}}(t),v_{a,A}^{\textrm{post}}(t)\bigr),$
where 
\begin{align}
v_{a,A}^{\textrm{post}}(t) & =\Bigl(\frac{1}{V_{a}(t)}+\frac{1}{\tilde{\sigma}_{a}^{2}(t)}\Bigr)^{-1},\label{eq:V_A_post_var}\\
z_{a,A}^{\textrm{post}}(t) & =v_{a,A}^{\textrm{post}}(t)\Bigl(\frac{Z_{a}(t)}{V_{a}(t)}+\frac{\tilde{y}_{a}(t)}{\tilde{\sigma}_{a}^{2}(t)}\Bigr).\label{eq:Z_A_post_equ}
\end{align}
Substituting (\ref{eq:V_A_post_var}), (\ref{eq:Z_A_post_equ}) into
(\ref{eq:var_z_ext}), (\ref{eq:mean_z_ext}), we obtain $v_{a,A}^{\textrm{ext}}(t)=V_{a}(t)$
and $z_{a,A}^{\textrm{ext}}(t)=Z_{a}(t)$. In the derivation of AMP
\cite{rangan2011generalized,krzakala2012probabilistic}, $V_{a}(t)$
and $Z_{a}(t)$ are computed as (D5) and (D6) of Algorithm \ref{CAMP}
in large system limit, which completes the proof.
\end{IEEEproof}

According to Lemma 1 and the framework in Algorithm \ref{Algorithm_GLM_framework},
we readily obtain a generalized version of AMP called Gr-AMP, as shown
in Algorithm \ref{G-AMP}. 

\begin{algorithm}
\protect\caption{Generalized version of AMP (Gr-AMP)}

\begin{raggedright}
1) Initialization: $\mathbf{z}_{A}^{\textrm{ext}}(0),\mathbf{v}_{A}^{\textrm{ext}}(0)$,
$t=1$;
\par\end{raggedright}

2) Update $\tilde{\boldsymbol{\sigma}}^{2}(t)$, $\mathbf{\tilde{y}}(t)$
as (\ref{eq:var_z_ext_B2A}) and (\ref{eq:mean_z_ext_B2A});

3) Perform AMP with $\textrm{Iter}{}_{SLM}$ iterations over $\mathbf{\tilde{y}}(t)=\mathbf{Ax}+\mathbf{\tilde{w}}(t)$,
$\mathbf{\tilde{w}}(t)\sim\mathcal{N}\bigl(\mathbf{\tilde{w}}(t);\mathbf{0},\mathsf{diag}(\tilde{\boldsymbol{\sigma}}^{2}(t))\bigr)$
and update $v_{a,A}^{\textrm{ext}}(t)=V_{a}(t)$, $z_{a,A}^{\textrm{ext}}(t)=Z_{a}(t)$
with the results of AMP. 

4) Set $t\leftarrow t+1$ and proceed to step 2) until $t>T_{\textrm{max}}$.
\label{G-AMP}
\end{algorithm}

Note that the proposed Gr-AMP in Algorithm \ref{G-AMP} can be viewed
as a general form of the well-known GAMP \cite{rangan2011generalized}.
In particular, if $\textrm{Iter}{}_{SLM}=1$, Gr-AMP becomes equivalent
to the original GAMP. To prove this, in step 2) of Algorithm \ref{G-AMP},
we first obtain the equivalent noise variance $\tilde{\sigma}_{a}^{2}(t)$
and observation $\tilde{y}_{a}(t)$ from (\ref{eq:var_z_ext_B2A})
and (\ref{eq:mean_z_ext_B2A}) with $v_{a,A}^{\textrm{ext}}(t-1)=V_{a}(t-1)$
and $z_{a,A}^{\textrm{ext}}(t-1)=Z_{a}(t-1)$. Then, in step 3) of
Algorithm \ref{G-AMP}, one iteration of AMP is performed by simply
replacing $\sigma^{2}$, $y_{a}$ in Algorithm \ref{CAMP} with $\tilde{\sigma}_{a}^{2}(t)$,
$\tilde{y}_{a}(t)$. After some algebra and the definitions of two
intermediate variables $\hat{s}_{a}(t-1)$ and $\tau_{a}^{s}(t-1)$
in (\ref{eq:sa_t_def}) and (\ref{eq:kesi_at_def}), the $t$-th iteration
of the proposed Gr-AMP in Algorithm \ref{G-AMP} with $\textrm{Iter}{}_{SLM}=1$
becomes
\begin{align}
\hat{s}_{a}(t-1) & =\frac{z_{a,B}^{\textrm{post}}(t)-Z_{a}(t-1)}{V_{a}(t-1)},\label{eq:sa_t_def}\\
\tau_{a}^{s}(t-1) & =\frac{V_{a}(t-1)-v_{a,B}^{\textrm{post}}(t)}{V_{a}^{2}(t-1)},\label{eq:kesi_at_def}
\end{align}

\begin{align}
\Sigma_{i}(t-1) & =\Bigl[\sum_{a}A_{ai}^{2}\tau_{a}^{s}(t-1)\Bigr]^{-1},\\
r_{i}(t-1) & =\hat{x}_{i}(t-1)+\Sigma_{i}(t-1)\sum_{a}A_{ai}\hat{s}_{a}(t-1),\label{eq:Rit_update}\\
\hat{x}_{i}(t) & =\mathsf{E}(x_{i}|r_{i}(t-1),\Sigma_{i}(t-1)),\label{eq:Za_t_update}\\
\tau_{i}(t) & =\mathsf{Var}(x_{i}|r_{i}(t-1),\Sigma_{i}(t-1)),\\
V_{a}(t) & =\sum_{i}A_{ai}^{2}\tau_{i}(t),\\
Z_{a}(t) & =\sum_{i}A_{ai}\hat{x}_{i}(t)-V_{a}(t)\hat{s}_{a}(t-1).\label{eq:Za_t_GAMP}
\end{align}
Recalling the definitions (\ref{eq:Z_post_B}) and (\ref{eq:Vz_post_B}),
it can be seen that Gr-AMP with $\textrm{Iter}{}_{SLM}=1$ is equivalent
to the original GAMP \cite{rangan2011generalized,Rangan2012GAMP}.
As a result, the proposed framework provides a new perspective on
GAMP and leads to a concise derivation.

\subsection{From VAMP and SBL to Gr-VAMP and Gr-SBL}

For VAMP and SBL, the values of $\mathbf{z}_{A}^{\textrm{ext}}$ and
$\mathbf{v}_{A}^{\textrm{ext}}$ are not already calculated as in
AMP. However, the posterior mean and covariance matrix of $\mathbf{x}$,
denoted as $\mathbf{\hat{x}}_{2k}$ and $\mathbf{C}_{2k}$, respectively,
can be obtained in the linear MMSE (LMMSE) operation of VAMP and SBL
(refer to \textcolor{black}{\cite{Rangan2016Vector}, \cite{tipping2001sparse},
and \cite{wipf2004sparse}}). Thus, the posterior mean and variance
of $\mathbf{z}$ are calculated as $\mathbf{z}_{A}^{\textrm{post}}=\mathbf{A}\mathbf{\hat{x}}_{2k}$,
$\mathbf{v}_{A}^{\textrm{post}}=\frac{1}{M}\mathsf{trace}(\mathbf{A}\mathbf{C}_{2k}\mathbf{A}^{T})$,
where $\mathsf{trace}(\mathbf{X})$ denotes the trace of matrix $\mathbf{X}$.
Then, we obtain $\mathbf{z}_{A}^{\textrm{ext}}$ and variance $\mathbf{v}_{A}^{\textrm{ext}}$
from (\ref{eq:var_z_ext}) and (\ref{eq:mean_z_ext}). The resulting
generalized VAMP (Gr-VAMP) and generalized SBL (Gr-SBL) are shown
in Algorithm \ref{Gr_VAMP_SBL}.

\begin{algorithm}
\protect\caption{Generalized VAMP/SBL (Gr-VAMP/Gr-SBL) }

\begin{raggedright}
1) Initialization: $\mathbf{z}_{A}^{\textrm{ext}}(0),\mathbf{v}_{A}^{\textrm{ext}}(0)$,
$t=1$;
\par\end{raggedright}

2) Update $\tilde{\boldsymbol{\sigma}}^{2}(t)$, $\mathbf{\tilde{y}}(t)$
as (\ref{eq:var_z_ext_B2A}) and (\ref{eq:mean_z_ext_B2A});

3) Perform VAMP/SBL with $\textrm{Iter}{}_{SLM}$ iterations over
$\mathbf{\tilde{y}}(t)=\mathbf{Ax}+\mathbf{\tilde{w}}(t)$, $\mathbf{\tilde{w}}(t)\sim\mathcal{N}\bigl(\mathbf{\tilde{w}}(t);\mathbf{0},\mathsf{diag}(\tilde{\boldsymbol{\sigma}}^{2}(t))\bigr)$;

4) Update $\mathbf{v}_{A}^{\textrm{ext}}(t)$, $\mathbf{z}_{A}^{\textrm{ext}}(t)$
as (\ref{eq:var_z_ext}) and (\ref{eq:mean_z_ext}) with $\mathbf{z}_{A}^{\textrm{post}}(t)=\mathbf{A}\mathbf{\hat{x}}_{2k}(t)$,
$\mathbf{v}_{A}^{\textrm{post}}(t)=\frac{1}{M}\mathsf{trace}(\mathbf{A}\mathbf{C}_{2k}(t)\mathbf{A}^{T})$,
where $\mathbf{\hat{x}}_{2k}(t)$ and $\mathbf{C}_{2k}(t)$ are the
posterior mean and covariance matrix obtained in the LMMSE step of
VAMP\textcolor{black}{/SBL,} respectively. 

5) Set $t\leftarrow t+1$ and proceed to step 2) until $t>T_{\textrm{max}}$.
\label{Gr_VAMP_SBL}
\end{algorithm}

\subsection{Applications to 1-bit Compressed Sensing}

In this subsection, we evaluate the performances of Gr-AMP, Gr-VAMP,
and Gr-SBL for 1-bit CS. Assume that the sparse signal $\mathbf{x}$
follows i.i.d. Bernoulli-Gaussian distribution $p_{0}\bigl(\mathbf{x}\bigr)=\prod_{i}\bigl(1-\rho\bigr)\delta\bigl(x_{i}\bigr)+\rho\mathcal{N}\bigl(x_{i};0,\rho^{-1}\bigr)$.
The 1-bit quantized measurements are obtained by $\mathbf{y}=\mathtt{\mathsf{sign}}\left(\mathbf{Ax}+\mathbf{w}\right)$,
where $\mathtt{\mathsf{sign}}(\cdot)$ is the component-wise sign
of each element. \textcolor{black}{We consider the case $N=512,M=2048$
for sparse ratio $\rho=0.1$}. The signal to noise ratio (SNR) $\mathsf{E}(\left\Vert \mathbf{Ax}\right\Vert _{2}^{2})/\mathsf{E}(\left\Vert \mathbf{w}\right\Vert _{2}^{2})$
is set to be 50 dB. \textcolor{black}{To test the performances for
general} matri\textcolor{black}{x $\mathbf{A}$, ill-conditioned }matrices\textcolor{black}{{}
with various condition numbers are considered following \cite{Rangan2016Vector}.
}Specifically, $\mathbf{A}$ is constructed from the singular value
decomposition (SVD) $\mathbf{A}=\mathbf{US}\mathbf{V}^{T}$, where
orthogonal matrices $\mathbf{U}$ and $\mathbf{V}$ were uniformly
drawn w.r.t. Harr measure. The singular values were chosen to achieve
a desired condition number $\kappa(\mathbf{A})=\lambda_{\textrm{max}}/\lambda_{\textrm{min}}$,
where $\lambda_{\textrm{max}}$ and $\lambda_{\textrm{min}}$ are
the maximum and minimum singular values of $\mathbf{A}$, respectively.\textcolor{red}{{}
}As in\textcolor{black}{{} \cite{Schniter2016Vector},} the recovery
performance is assessed using the debiased normalized mean square
error (dNMSE) in decibels, i.e.,$\underset{c}{\textrm{min}}\:20\log_{10}(\left\Vert c\mathbf{\hat{x}}-\mathbf{x}\right\Vert _{2}/\left\Vert \mathbf{x}\right\Vert _{2})$.
In all simulations, $\textrm{Iter}{}_{SLM}=1,T_{\textrm{max}}=50,\mathbf{z}_{A}^{\textrm{ext}}(0)=0,\mathbf{v}_{A}^{\textrm{ext}}(0)=10^{8}$
and the results are averaged over 100 realizations. The results of
GAMP and the algorithm in\textcolor{black}{{} \cite{Schniter2016Vector},
denoted as GVAMP, are also given for comparison. }

\begin{figure}
\begin{centering}
\includegraphics[width=9.5cm]{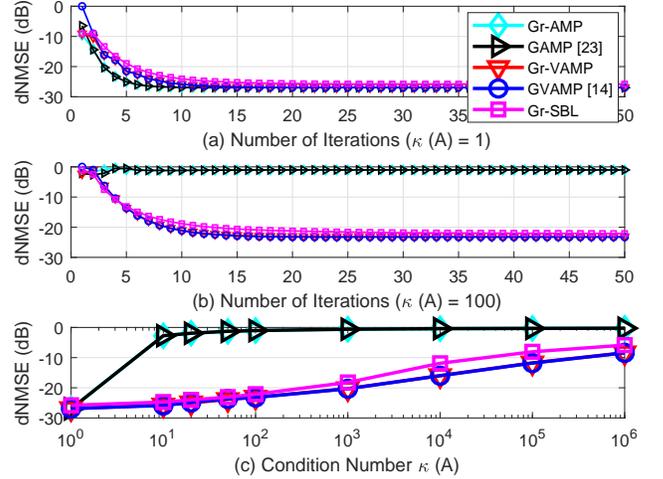}
\par\end{centering}

\protect\caption{dNMSE versus number of iterations for (a): $\kappa(\mathbf{A})$=
1 and (b): $\kappa(\mathbf{A})$= 100; (c) dNMSE versus the condition
number $\kappa(\mathbf{A})$. }
\label{fig:MSE_performance} 
\end{figure}

Fig. \ref{fig:MSE_performance} (a) and (b) show the results of dNMSE
versus the number of algorithm iterations when the condition number
$\kappa(\mathbf{A})=1\textrm{ and }100$, respectively. When $\kappa(\mathbf{A})=1$,
all algorithms converge to the same dNMSE value except Gr-SBL which
has slightly worse performance. When $\kappa(\mathbf{A})=100$, however,
both Gr-AMP and GAMP fail while other algorithms still work well.
\textcolor{black}{In }Fig. \ref{fig:MSE_performance} (c), the dNMSE
is evaluated for various condition number $\kappa(\mathbf{A})$ ranging
from 1 (i.e., row-orthogonal matrix $\mathbf{A}$) to $1\times10^{6}$
(i.e., highly ill-conditioned matrix $\mathbf{A}$). It is seen that
as the condition number $\kappa(\mathbf{A})$ increases, the recovery
performances degrade smoothly for Gr-VAMP, GVAMP, and Gr-SBL while
both Gr-AMP and GAMP diverge for even mild condition number. 

\section{Conclusion}

In this letter a unified Bayesian inference framework for GLM is presented
whereby a variety of SLM algorithms can be easily extended to handle
the GLM problem. Specific instances elucidated under this framework
are the extensions of AMP, VAMP, and SBL, which unveils intimate connections
between some established algorithms and leads to novel extensions
for some other SLM algorithms. Numerical results for 1-bit CS demonstrate
the effectiveness of this framework.

\section*{Acknowledgments }

The authors sincerely thank P. Schniter for valuable discussions and
the anonymous reviewers for valuable comments.

\newpage{}

\appendix{}

\subsection{Details of Gr-VAMP}

The vector approximate message passing (VAMP) algorithm\textcolor{black}{\cite{Rangan2016Vector}}
is a robust version of the well-known AMP for standard linear model
(SLM). Define the LMMSE estimator $g_{2}(\mathbf{r}_{2k},\gamma_{2k})$
and component-wise denoiser function $\boldsymbol{g}_{1}(\mathbf{r}_{1k},\gamma_{1k})$
as
\begin{align}
g_{2}(\mathbf{r}_{2k},\gamma_{2k}) & =\Bigl(\frac{\mathbf{A}^{T}\mathbf{A}}{\sigma^{2}}+\gamma_{2k}\mathbf{I}_{N}\Bigr)^{-1}\Bigl(\frac{\mathbf{A}^{T}\mathbf{y}}{\sigma^{2}}+\gamma_{2k}\mathbf{r}_{2k}\Bigr),\label{eq:g2}\\
g_{1}(r_{1k},\gamma_{1k}) & =\mathsf{E}(x_{i}|r_{1k},\gamma_{1k}),\label{eq:g1}
\end{align}
where the expectation in calculating the posterior mean $\mathsf{E}(x_{i}|r_{1k},\gamma_{1k})$
is with respect to (w.r.t.) the distribution $q_{x}^{t}\left(x_{i}\right)\propto p_{0}(x_{i})\mathcal{N}\bigl(x_{i};r_{1k},1/\gamma_{1k}\bigr).$ 

Then, the VAMP algorithm is shown in Algorithm \ref{VAMP}. Note that
here we only show the MMSE version of VAMP. However, it also applies
to the SVD version of VAMP. For more details, please refer to \textcolor{black}{\cite{Rangan2016Vector}.
It should also be noted that the denoising step and the LMMSE step
in }Algorithm \ref{VAMP}\textcolor{black}{{} can be exchanged in practical
implementations. }

\begin{algorithm}
\protect\caption{VAMP Algorithm\textcolor{black}{\cite{Rangan2016Vector}}}

\begin{raggedright}
\textbf{Require}: LMMSE estimator $g_{2}(\mathbf{r}_{2k},\gamma_{2k})$,
denoiser $g_{1}(\mathbf{r}_{1k},\gamma_{1k})$, number of iterations
$\textrm{Iter}{}_{SLM}$
\par\end{raggedright}

\begin{raggedright}
1: Initialization: $\mathbf{r}_{10}$ and $\gamma_{10}\geq0$
\par\end{raggedright}

2: \textbf{For} $k=1,2,...,\textrm{Iter}{}_{SLM}$, \textbf{Do}

3: $\quad$//Denoising 

4: $\quad$$\mathbf{\widehat{x}}_{1k}=\boldsymbol{g}_{1}(\mathbf{r}_{1k},\gamma_{1k})$

5: $\quad$$\mathbf{\alpha}_{1k}=\bigl\langle\boldsymbol{g}_{1}^{\prime}(\mathbf{r}_{1k},\gamma_{1k})\bigr\rangle$

6: $\quad$$\mathbf{\eta}_{1k}=\gamma_{1k}/\mathbf{\alpha}_{1k}$

7: $\quad$$\mathbf{\gamma}_{2k}=\mathbf{\eta}_{1k}-\gamma_{1k}$

8: $\quad$$\mathbf{r}_{2k}=\bigl(\mathbf{\eta}_{1k}\mathbf{\widehat{x}}_{1k}-\gamma_{1k}\mathbf{r}_{1k}\bigr)/\mathbf{\gamma}_{2k}$ 

9: 

10: $\quad$//LMMSE

11: $\quad$$\mathbf{\widehat{x}}_{2k}=\boldsymbol{g}_{2}(\mathbf{r}_{2k},\gamma_{2k})$

12: $\quad$$\mathbf{\alpha}_{2k}=\bigl\langle\boldsymbol{g}_{2}^{\prime}(\mathbf{r}_{2k},\gamma_{2k})\bigr\rangle$

13: $\quad$$\mathbf{\eta}_{2k}=\gamma_{2k}/\mathbf{\alpha}_{2k}$

14: $\quad$$\mathbf{\gamma}_{1k+1}=\mathbf{\eta}_{2k}-\gamma_{2k}$

15: $\quad$$\mathbf{r}_{1k+1}=\bigl(\mathbf{\eta}_{2k}\mathbf{\widehat{x}}_{2k}-\gamma_{2k}\mathbf{r}_{2k}\bigr)/\mathbf{\gamma}_{1k+1}$ 

16: \textbf{end for}

17: Return $\mathbf{\widehat{x}}_{1k}$

\label{VAMP}
\end{algorithm}

Note that line 12 of Algorithm \ref{VAMP} represents
\begin{equation}
\bigl\langle\boldsymbol{g}_{2}^{\prime}(\mathbf{r}_{2k},\gamma_{2k})\bigr\rangle=\frac{\gamma_{2k}}{N}\mathsf{trace}(\Bigl(\frac{\mathbf{A}^{T}\mathbf{A}}{\sigma^{2}}+\gamma_{2k}\mathbf{I}_{N}\Bigr)^{-1}),\label{eq:g2_deriv}
\end{equation}
where $\mathsf{trace}(\mathbf{X})$ denotes the trace of matrix $\mathbf{X}$.
Define
\begin{equation}
\mathbf{C}_{2k}=\Bigl(\frac{\mathbf{A}^{T}\mathbf{A}}{\sigma^{2}}+\gamma_{2k}\mathbf{I}_{N}\Bigr)^{-1},\label{eq:C2k_def}
\end{equation}
then $\bigl\langle\boldsymbol{g}_{2}^{\prime}(\mathbf{r}_{2k},\gamma_{2k})\bigr\rangle=\frac{\gamma_{2k}}{N}\mathsf{trace}(\mathbf{C}_{2k})$. 

As shown in \textcolor{black}{\cite{Rangan2016Vector}, }line 11 and
12 of Algorithm \ref{VAMP} can be recognized as the MMSE estimate
of $\mathbf{x}$ under likelihood $\mathcal{N}\bigl(\mathbf{y};\mathbf{Ax},\sigma^{2}\mathbf{I}_{M}\bigr)$
and prior $\mathbf{x}\sim\mathcal{N}\bigl(\mathbf{x};\mathbf{r}_{2k},\gamma_{2k}^{-1}\bigr)$.
Specifically, the posterior mean and covariance matrix are $\mathbf{\widehat{x}}_{2k}$
and $\mathbf{C}_{2k}$, respectively. 

As shown in the unified framework in Algorithm \ref{Algorithm_GLM_framework},
the key lies in calculating the extrinsic values of $\mathbf{z}_{A}^{\textrm{ext}}$
and $\mathbf{v}_{A}^{\textrm{ext}}$. To this end, from (\ref{eq:var_z_ext})
and (\ref{eq:mean_z_ext}), we should first calculate the posterior
mean vector $\mathbf{z}_{A}^{\textrm{post}}$ and variance vector
$\mathbf{v}_{A}^{\textrm{post}}$ of $\mathrm{\mathbf{z}}$ from the
results of VAMP. As shown in Algorithm \ref{VAMP}, the posterior
mean $\mathbf{\widehat{x}}_{2k}$ and covariance matrix $\mathbf{C}_{2k}$
of $\mathbf{x}$ can be obtained via the LMMSE step of VAMP. Since
$\mathbf{z}=\mathbf{Ax}$, the posterior mean and covariance matrix
of $\mathbf{z}$ are
\begin{align}
\mathbf{z}_{A}^{\textrm{post}} & =\mathbf{A}\mathbf{\widehat{x}}_{2k},\label{eq:z_post}\\
\mathbf{C}_{z} & =\mathbf{A}\mathbf{C}_{2k}\mathbf{A}^{T}.\label{eq:z_C_mat}
\end{align}

As a result, variance vector $\mathbf{v}_{A}^{\textrm{post}}$ of
$\mathrm{\mathbf{z}}$ can be calculated as the diagonal vector of
covariance matrix $\mathbf{C}_{z}$, i.e.,
\begin{equation}
\mathbf{v}_{A}^{\textrm{post}}=\mathsf{diag}(\mathbf{A}\mathbf{C}_{2k}\mathbf{A}^{T}).\label{eq:z_var_vector}
\end{equation}

Note that for AMP, the covariance matrix of $\mathbf{x}$ is unavailable,
so that we could not directly calculate $\mathbf{z}_{A}^{\textrm{post}}$
and $\mathbf{v}_{A}^{\textrm{post}}$ as (\ref{eq:z_post}) and (\ref{eq:z_var_vector}),
respectively. In fact, only the diagonal elements of the covariance
matrix of $\mathbf{x}$ are available in AMP. 

Moreover, in VAMP\textcolor{black}{\cite{Rangan2016Vector}}, the
variances values are averaged before being further processed. To be
consistent with the implementation of VAMP, the variance vector $\mathbf{v}_{A}^{\textrm{post}}$
of $\mathrm{\mathbf{z}}$ are also set to be the average value of
the diagonal elements of covariance matrix $\mathbf{C}_{z}$, i.e.,
\begin{equation}
\mathbf{v}_{A}^{\textrm{post}}=\frac{1}{M}\mathsf{trace}(\mathbf{A}\mathbf{C}_{2k}\mathbf{A}^{T}).\label{eq:v_post}
\end{equation}

Given $\mathbf{z}_{A}^{\textrm{post}}$ and $\mathbf{v}_{A}^{\textrm{post}}$,
we can then obtain $\mathbf{z}_{A}^{\textrm{ext}}$ and variance $\mathbf{v}_{A}^{\textrm{ext}}$
from (10) and (11). Finally, from the unified algorithm framework
in Algorithm \ref{Algorithm_GLM_framework}, we obtain the generalized
VAMP (Gr-VAMP) as shown in Algorithm \ref{Gr_VAMP_SBL}.

\subsection{Details of Gr-SBL}

The sparse Bayesian learning (SBL) algorithm\textcolor{black}{\cite{tipping2001sparse}\cite{wipf2004sparse}}
is a well-known Bayesian inference method for compressed sensing for
SLM (\ref{eq:linear_model}). In SBL, the prior distribution of signal
$\mathbf{x}$ is assumed to be i.i.d Gaussian, i.e.,
\begin{equation}
p_{0}\bigl(\mathbf{x}|\boldsymbol{\alpha}\bigr)=\prod_{i}\mathcal{N}\bigl(x_{i};0,\alpha_{i}^{-1}\bigr),\label{eq:pri_SBL}
\end{equation}
where $\boldsymbol{\alpha}\triangleq\left\{ \alpha_{i}\right\} $
are non-negative hyper-parameters controlling the sparsity of the
signal $\mathbf{x}$ and they follow Gamma distributions 
\begin{equation}
p_{0}\bigl(\boldsymbol{\alpha}\bigr)=\prod_{i}\mathsf{Gamma}\bigl(x_{i}|a,b\bigr)=\prod_{i}\mathsf{\varGamma}\bigl(a\bigr)^{-1}b^{a}\alpha_{i}^{a}e^{-b\alpha_{i}},
\end{equation}
where $\mathsf{\varGamma}\bigl(a\bigr)$ is the Gamma function. 

Then, using the expectation maximization (EM) method, the conventional
SBL algorithm is shown in Algorithm \ref{SBL}. Here we assumed knowledge
of noise variance. In fact, SBL can also handle the case with unknown
noise variance. For more details, please refer to \textcolor{black}{\cite{tipping2001sparse}\cite{wipf2004sparse}.}

\begin{algorithm}
\protect\caption{SBL Algorithm\textcolor{black}{\cite{tipping2001sparse}\cite{wipf2004sparse}}}

\begin{raggedright}
\textbf{Require}: Set the values of parameters $a,b$, number of iterations
$\textrm{Iter}{}_{SLM}$.
\par\end{raggedright}

\begin{raggedright}
1: Initialization: $\alpha_{i},i=1,...,N$.
\par\end{raggedright}

2: \textbf{For} $k=1,2,...,\textrm{Iter}{}_{SLM}$, \textbf{Do}

3: $\quad$//LMMSE

4: $\quad$$\mathbf{C}_{2k}=\Bigl(\frac{\mathbf{A}^{T}\mathbf{A}}{\sigma^{2}}+\mathsf{diag}\bigl(\boldsymbol{\alpha}\bigr)\Bigr)^{-1}$

5: $\quad$$\mathbf{\widehat{x}}_{2k}=\mathbf{C}_{2k}\frac{\mathbf{A}^{T}\mathbf{y}}{\sigma^{2}}$

6: 

7: $\quad$//Parameters updating

8: $\quad$$\alpha_{i}=\frac{1+2a}{\mathbf{\widehat{x}}_{2k,i}^{2}+\Sigma_{ii}+2b}$,
where $\Sigma_{ii}$ is the $i$-th diagonal 

9: $\quad$element of $\mathbf{C}_{2k}$, $\mathbf{\widehat{x}}_{2k,i}$
is the $i$-th element of $\mathbf{\widehat{x}}_{2k}$.

10: \textbf{end for}

11: Return $\mathbf{\widehat{x}}_{2k}$

\label{SBL}
\end{algorithm}

It is seen that in SBL, line 4 and 5 of Algorithm \ref{SBL} can be
recognized as the LMMSE estimate of $\mathbf{x}$ under likelihood
$\mathcal{N}\bigl(\mathbf{y};\mathbf{Ax},\sigma^{2}\mathbf{I}_{M}\bigr)$
and Gaussian prior $\mathbf{x}\sim p_{0}\bigl(\mathbf{x}|\boldsymbol{\alpha}\bigr)$
given $\boldsymbol{\alpha}$. Thus, the posterior mean and covariance
matrix of $\mathbf{x}$ can be calculated as $\mathbf{\widehat{x}}_{2k}$
and $\mathbf{C}_{2k}$ in Algorithm \ref{SBL}, respectively. 

Under the unified framework in Algorithm \ref{Algorithm_GLM_framework},
and similar to the derivation of Gr-VAMP, we first obtain the posterior
mean and covariance matrix of $\mathbf{z}$ as
\begin{align}
\mathbf{z}_{A}^{\textrm{post}} & =\mathbf{A}\mathbf{\widehat{x}}_{2k},\label{eq:z_post-1}\\
\mathbf{C}_{z} & =\mathbf{A}\mathbf{C}_{2k}\mathbf{A}^{T}.\label{eq:z_C_mat-1}
\end{align}

Then, the variance vector $\mathbf{v}_{A}^{\textrm{post}}$ of $\mathrm{\mathbf{z}}$
can be calculated to be the average of the diagonal vector of covariance
matrix $\mathbf{C}_{z}$, i.e.,
\begin{equation}
\mathbf{v}_{A}^{\textrm{post}}=\frac{1}{M}\mathsf{trace}(\mathbf{A}\mathbf{C}_{2k}\mathbf{A}^{T}).\label{eq:z_var_vector-1}
\end{equation}

Note that the average in (\ref{eq:z_var_vector-1}) is not essential
and we can also simply choose the diagonal elements of $\mathbf{C}_{z}$,
i.e.,
\begin{equation}
\mathbf{v}_{A}^{\textrm{post}}=\mathsf{diag}(\mathbf{A}\mathbf{C}_{2k}\mathbf{A}^{T}).
\end{equation}

Given $\mathbf{z}_{A}^{\textrm{post}}$ and $\mathbf{v}_{A}^{\textrm{post}}$,
we obtain $\mathbf{z}_{A}^{\textrm{ext}}$ and variance $\mathbf{v}_{A}^{\textrm{ext}}$
from (\ref{eq:var_z_ext}) and (\ref{eq:mean_z_ext}). Finally, from
the unified algorithm framework in Algorithm \ref{Algorithm_GLM_framework},
we obtain the generalized SBL (Gr-SBL) as shown in Algorithm \ref{Gr_VAMP_SBL}.

\bibliographystyle{IEEEtran}
\bibliography{IEEEabrv,Mybib}

\end{document}